# BioimageAIpub: a toolbox for AI-ready bioimaging data publishing

Stefan Dvoretskii[1,*], Anwai Archit[2,*,#], Constantin Pape[2,3,4], Josh Moore[5,6,7], Marco Nolden[1,8,9]

[1] DKFZ Division of Medical Computing

[2] Georg-August-University Göttingen, Institute of Computer Science

[3] CAIMed - Lower Saxony Center for AI & Causal Methods in Medicine, Göttingen

[4] Cluster of Excellence Multiscale Bioimaging (MBExC), Georg-August-University Göttingen

[5] German BioImaging – Gesellschaft für Mikroskopie und Bildanalyse e.V., Konstanz, Germany

[6] National Research Data Infrastructure for Microscopy and BioImage Analysis (NFDI4BIOIMAGE)

[7] Open Microscopy Environment (OME)

[8] Helmholtz Metadata Collaboration (HMC) Hub Health, DKFZ, Heidelberg, Germany

[9] Pattern Analysis and Learning Group, Heidelberg University Hospital

[*] contributed equally, [#] corresponding author

# Abstract

Modern bioimage analysis approaches are data hungry, making it necessary for researchers to scavenge data beyond those collected within their (bio)imaging facilities. In addition to scale, bioimaging datasets must be accompanied with suitable, high-quality annotations and metadata. Although established data repositories such as the Image Data Resource (IDR) and BioImage Archive offer rich metadata, their contents typically cannot be directly consumed by image analysis tools without substantial data wrangling. Such a tedious assembly and conversion of (meta)data can account for a dedicated amount of time investment for researchers, hindering the development of more powerful analysis tools. Here, we introduce BioimageAIpub, a workflow that streamlines bioimaging data conversion, enabling a seamless upload to HuggingFace, a widely used platform for sharing machine learning datasets and models.

The rise of bioimaging data is enabling unprecedented advances in the field of artificial intelligence (AI) in bioimaging [1], including multimodal learning [2] and foundation models for bioimaging [3]. While these approaches open new avenues of biological discovery, they are extremely data hungry, making it necessary for researchers to leverage datasets beyond those imaged in home institutions or consortia [4]. In addition to scale, the bioimaging field in particular demands abundant, expert-level annotations and high-quality metadata, to ensure that datasets are interpretable, reusable, and suitable for downstream machine learning tasks. Popular data repositories like Image Data Resource (IDR) [5] and BioImage Archive [6], despite offering a rich assortment of metadata, are not readily interoperable with modern AI frameworks. As a result, researchers often must perform extensive data wrangling, i.e. assembling, harmonizing, and converting both images and corresponding metadata. Such tasks can contribute to consuming 70-80% for the development and benchmarking of state-of-the-art AI frameworks [7].

We present **BioimageAIpub**, a workflow that streamlines bioimaging data conversion and uploads to HuggingFace, an open-source platform considered as the "GitHub for Machine Learning". HuggingFace's Datasets API enables direct loading of datasets into popular AI frameworks like JAX, PyTorch and Tensorflow [8], eliminating the need for extensive data wrangling before training or evaluating AI models. Furthermore, CroissantML [9] and Xetare, which are supported within the HuggingFace ecosystem, capture key dataset metadata such as license, keywords, and annotated variables that can be processed in machine-actionable fashion. This reduces the manual effort to search relevant data and helps users to quickly assess whether an imaging dataset is suitable for their analysis task[10]. BioimageAIpub puts special emphasis on preserving and exposing image-associated annotations and metadata. These elements are critical for downstream biological applications; recreating them incurs a substantial cost [3]. BioimageAIpub complements existing tools such as MDEmic for metadata annotation [11], BioimageIT for analysis and methods integration [12], and MethodsJ2 for generating microscopy analysis workflows [13]. Compared to these efforts, our tool focuses specifically on making (annotated) bioimaging data and metadata easily publishable, and readily reusable in modern AI pipelines in a couple of seconds. We used BioimageAIpub to publish a RNAi knockdown bioimaging dataset from the IDR to HuggingFace (https://huggingface.co/datasets/stefanches/genomic-bioimaging). Despite its relatively specific focus, the dataset already received more than 1,200 downloads per month. We anticipate an even higher adoption in future, particularly for datasets that align with widely studied benchmarks and common model-development workflows [3].

BioimageAIpub collects data either from local storage or from an Amazon S3 bucket. For S3 sources, it can inspect bucket contents to support partial dataset operations in low-resource settings. After the file collection stage, BioimageAIpub handles conversion of the raw data to any standard imaging file formats. The current recommendation is to store data into 16-bit PNG and TIFF formats for the image data to reduce storage restrictions, since datasets over 1TB on HuggingFace currently require a special written justification. Additionally, BioimageAIpub can fetch metadata from OMERO using OMExcavator [14] and from IDR annotations. Users can also associate additional annotations with their data annotation directly. BioimageAIpub aims to support further metadata sources in the future. The resulting metadata is stored for each dataset split as a CSV file, which enables the Dataset Preview feature on HuggingFace. Dataset

Preview allows users to browse the annotations and run SQL queries directly in the browser before downloading the large bioimaging dataset.

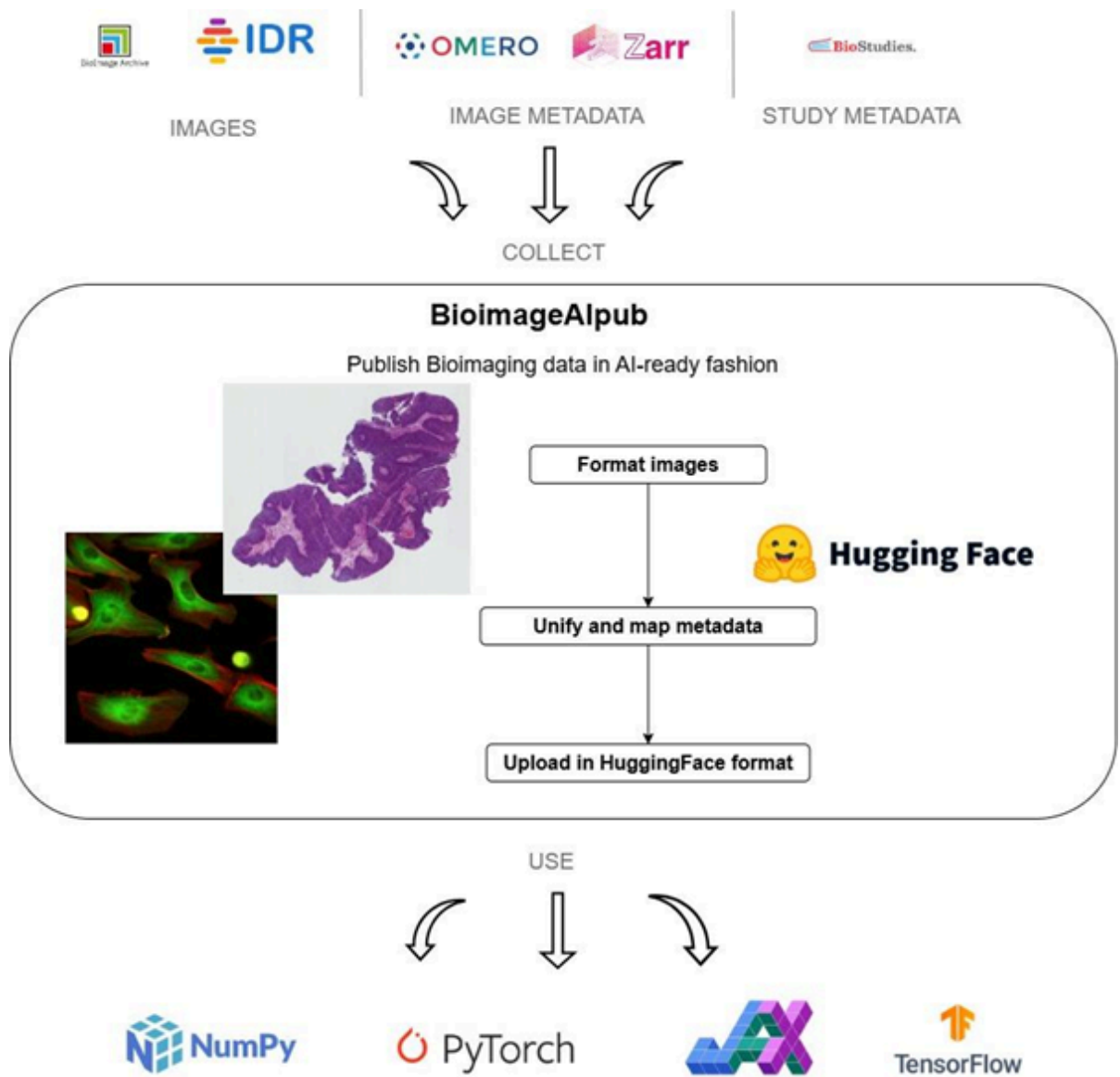

<u>Figure 1:</u> Overview of BioimageAIpub, a workflow that streamlines the conversion and publication of bioimaging datasets, together with their paired annotations. The pipeline also consolidates high-quality metadata to enable seamless upload to HuggingFace. After publication, users can access the dataset through CroissantML's programmatic interface and obtain tensors that are ready for downstream bioimaging AI pipelines.

Dataset-level annotations are assembled from multiple sources, including the IDR, BioImage Archive, and BioStudies [15]. Users are prompted to provide essential dataset metadata such as licensing, authorship and citations in the HuggingFace Dataset Card. Currently, this step is performed manually. Once all components are assembled, BioimageAIpub uploads the data and metadata to a HuggingFace repository, where they become immediately available for downstream tools and workflows.

In summary, BioimageAIpub enables users to publish high-quality annotated bioimaging data on HuggingFace. This makes bioimaging data readily usable with modern AI tooling. In addition, CroissantML and HuggingFace's Dataset Preview allow users to explore and analyze standardized metadata directly in the browser without downloading the full dataset. Thanks to the CroissantML and Dataset Preview features of HuggingFace, this also enables users to analyze standardized metadata without any data downloads.

Looking ahead, BioimageAIpub aims to extend automatic support for bioimaging file formats [16] and metadata sources [17, 18], as well as latest AI trends like automatic generation of dataset cards with large language models. Such functionality would require robust safeguards to ensure factual accuracy and prevent hallucinated content. Overall, BioimageAIpub aims to serve as a foundation for converting and publishing bioimaging datasets and publishing bioimaging datasets in an AI-ready format, enabling end-users to discover, assess and reuse both images and annotations to build the next generation bioimage analysis models.

### Code Availability

BioimageAIpub is open-source and available under Apache-2.0 license at https://github.com/German-BioImaging/bioimageaipub. The code is also available under Helmholtz Metadata Collaboration's GitLab (https://codebase.helmholtz.cloud/hmc-public/hmc-hub-health/bioimageaipub/) and on the EU Workflow Hub (https://workflowhub.eu/workflows/2034).

### Data Availability

The tooling was used to convert the IDR0012 study (https://idr.openmicroscopy.org/study/idr0012/) to HuggingFace. The resulting dataset is available at https://huggingface.co/datasets/stefanches/genomic-bioimaging.

**Acknowledgements**

This project was supported by the German Cancer Research Center (DKFZ), the Helmholtz Metadata Collaboration (HMC), an incubator-platform of the Helmholtz Association within the framework of the Information and Data Science strategic initiative. The work of Anwai Archit was funded by the Deutsche Forschungsgemeinschaft (DFG, German Research Foundation) - PA 4341/2-1. Constantin Pape is supported by the German Research Foundation (Deutsche Forschungsgemeinschaft, DFG) under Germany's Excellence Strategy – EXC 2067/1-390729940. This work is supported by the Ministry of Science and Culture of Lower Saxony through funds from the program zukunft.niedersachsen of the Volkswagen Foundation for the 'CAIMed – Lower Saxony Center for Artificial Intelligence and Causal Methods in Medicine' project (grant no. ZN4257). Josh Moore is supported by the Deutsche Forschungsgemeinschaft (DFG, German Research Foundation; 501864659 as part of NFDI4BIOIMAGE).